\documentclass[twocolumn,floatfix,superscriptaddress]{revtex4}
\usepackage{graphicx}
\usepackage{epstopdf}
\begin{document}
\newcommand\Hbar{$\overline{\rm H}$}
\newcommand\Hbars{$\overline{\rm H}$s}
\newcommand\pbar{$\overline{\rm p}$}
\newcommand\pos{e$^+$}
\newcommand\elec{e$^-$}
\newcommand\us{$\,$}

\title{A novel antiproton radial diagnostic based on octupole induced ballistic loss}
\author{G.B. Andresen}
\affiliation{Department of Physics and Astronomy, Aarhus University, DK-8000 Aarhus C, Denmark}
\author{W. Bertsche}
\affiliation{Department of Physics, Swansea University, Swansea SA2 8PP, United Kingdom}
\author{P.D. Bowe}
\affiliation{Department of Physics and Astronomy, Aarhus University, DK-8000 Aarhus C, Denmark}
\author{C.C. Bray}
\affiliation{Department of Physics, University of California at Berkeley, Berkeley, CA 94720-7300, USA}
\author{E. Butler}
\affiliation{Department of Physics, Swansea University, Swansea SA2 8PP, United Kingdom}
\author{C.L. Cesar}
\affiliation{Instituto de F\'{i}sica, Universidade Federal do Rio de Janeiro, Rio de Janeiro 21941-972, Brazil}
\author{S. Chapman}
\affiliation{Department of Physics, University of California at Berkeley, Berkeley, CA 94720-7300, USA}
\author{M. Charlton}
\affiliation{Department of Physics, Swansea University, Swansea SA2 8PP, United Kingdom}
\author{J. Fajans}
\affiliation{Department of Physics, University of California at Berkeley, Berkeley, CA 94720-7300, USA}
\author{M.C. Fujiwara}
\affiliation{TRIUMF, 4004 Wesbrook Mall, Vancouver BC, Canada V6T 2A3}
\author{R. Funakoshi}
\affiliation{Department of Physics, University of Tokyo, Tokyo 113-0033, Japan}
\author{D.R. Gill}
\affiliation{TRIUMF, 4004 Wesbrook Mall, Vancouver BC, Canada V6T 2A3}
\author{J.S. Hangst}
\affiliation{Department of Physics and Astronomy, Aarhus University, DK-8000 Aarhus C, Denmark}
\author{W.N. Hardy}
\affiliation{Department of Physics and Astronomy, University of British Columbia, Vancouver BC, Canada V6T 1Z4}
\author{R.S. Hayano}
\affiliation{Department of Physics, University of Tokyo, Tokyo 113-0033, Japan}
\author{M.E. Hayden}
\affiliation{Department of Physics, Simon Fraser University, Burnaby BC, Canada V5A 1S6}
\author{A.J. Humphries}
\affiliation{Department of Physics, Swansea University, Swansea SA2 8PP, United Kingdom}
\author{R. Hydomako}
\affiliation{Department of Physics and Astronomy, University of Calgary, Calgary AB, Canada T2N 1N4}
\author{M.J. Jenkins}
\affiliation{Department of Physics, Swansea University, Swansea SA2 8PP, United Kingdom}
\author{L.V. J\o rgensen}
\affiliation{Department of Physics, Swansea University, Swansea SA2 8PP, United Kingdom}
\author{L. Kurchaninov}
\affiliation{TRIUMF, 4004 Wesbrook Mall, Vancouver BC, Canada V6T 2A3}
\author{R. Lambo}
\affiliation{Instituto de F\'{i}sica, Universidade Federal do Rio de Janeiro, Rio de Janeiro 21941-972, Brazil}
\author{N. Madsen}
\affiliation{Department of Physics, Swansea University, Swansea SA2 8PP, United Kingdom}
\author{P. Nolan}
\affiliation{Department of Physics, University of Liverpool, Liverpool L69 7ZE, United Kingdom}
\author{K. Olchanski}
\affiliation{TRIUMF, 4004 Wesbrook Mall, Vancouver BC, Canada V6T 2A3}
\author{A. Olin}
\affiliation{TRIUMF, 4004 Wesbrook Mall, Vancouver BC, Canada V6T 2A3}
\author{R.D. Page}
\affiliation{Department of Physics, University of Liverpool, Liverpool L69 7ZE, United Kingdom}
\author{A. Povilus}
\affiliation{Department of Physics, University of California at Berkeley, Berkeley, CA 94720-7300, USA}
\author{P. Pusa}
\affiliation{Department of Physics, University of Liverpool, Liverpool L69 7ZE, United Kingdom}
\author{F. Robicheaux}
\affiliation{Department of Physics, Auburn University, Auburn, AL 36849-5311, USA}
\author{E. Sarid}
\affiliation{Department of Physics, NRCN-Nuclear Research Center Negev, Beer Sheva, IL-84190, Israel}
\author{S. Seif El Nasr}
\affiliation{Department of Physics and Astronomy, University of British Columbia, Vancouver BC, Canada V6T 1Z4}
\author{D.M. Silveira}
\affiliation{Instituto de F\'{i}sica, Universidade Federal do Rio de Janeiro, Rio de Janeiro 21941-972, Brazil}
\author{J.W. Storey}
\affiliation{TRIUMF, 4004 Wesbrook Mall, Vancouver BC, Canada V6T 2A3}
\author{R.I. Thompson}
\affiliation{Department of Physics and Astronomy, University of Calgary, Calgary AB, Canada T2N 1N4}
\author{D.P. van der Werf}
\affiliation{Department of Physics, Swansea University, Swansea SA2 8PP, United Kingdom}
\author{J.S. Wurtele}
\affiliation{Department of Physics, University of California at Berkeley, Berkeley, CA 94720-7300, USA}
\author{Y. Yamazaki}
\affiliation{Atomic Physics Laboratory, RIKEN, Saitama 351-0198, Japan}
\collaboration{ALPHA Collaboration}
\noaffiliation

\date{Received \today}

\begin{abstract} We report results from a novel diagnostic that probes the outer radial profile of trapped antiproton clouds.  The diagnostic allows us to determine the profile by monitoring the time-history of antiproton losses that occur as an octupole field in the antiproton confinement region is increased.  We show several examples of how this diagnostic helps us to understand the radial dynamics of antiprotons in normal and nested  Penning-Malmberg traps. Better understanding of these dynamics may aid current attempts to trap antihydrogen atoms.
\end{abstract}


\maketitle

\section{Introduction}

Cold antihydrogen atoms (\Hbar) were first produced by the ATHENA collaboration \cite{amor:02}, and, shortly thereafter, by ATRAP \cite{gabr:02} at the CERN Antiproton Decelerator (AD) \cite{maur:97} in 2002.  They were produced by mixing positrons (\pos) and antiprotons (\pbar) held in Penning-Malmberg traps.  Such traps use a solenoidal axial magnetic field $B_z$ to provide radial confinement, and electrostatic wells to provide axial confinement.  Penning-Malmberg traps confine only charged particles and, consequently, do not confine neutral \Hbar\ atoms.

The current generation of experiments \cite{andr:07,gabr:07} aims to trap  \Hbar\ atoms as this is likely necessary for precision CPT and gravity tests.  Neutral \Hbar\ atoms have a small permanent magnetic moment, and can be trapped in the magnetic minimum of a so-called Minimum-B trap \cite{prit:83}.  The magnetic minimum can be created by two axially separated mirror coils which create an axial minimum, and a multipole field, such as an octupole \cite{faja:04,bert:06}, which creates the radial minimum.  In all current schemes, the Minimum-B and Penning-Malmberg traps must be co-located because the \pbar's, \pos's, and \Hbars\ must all be trapped in the same spatial region.  Thus, in cylindrical coordinates $(r,\theta,z)$, the net magnetic field will be
\begin{equation}
\label{octupole}
\mathbf{B}= B_z{\hat z}+B_w{\left(\frac{r}{R_w}\right)}^3\left[{\hat r}\cos(4\theta)-{\hat \theta}\sin(4\theta)\right]+\mathbf{B}_{\rm M}(r,\theta,z)
\end{equation}
when using an octupole.  Here $R_w$ is the trap wall radius, $B_w$ is the octupole field at the wall, and $\mathbf{B}_{\rm M}(r,z)$ is the field of the mirror coils.  The mirror coils were not energized for the data taken for this paper; henceforth we will set $\mathbf{B}_{\rm M}=0$.

\begin{figure*}[t]
\centering
\centerline{\resizebox{16cm}{!}{\includegraphics{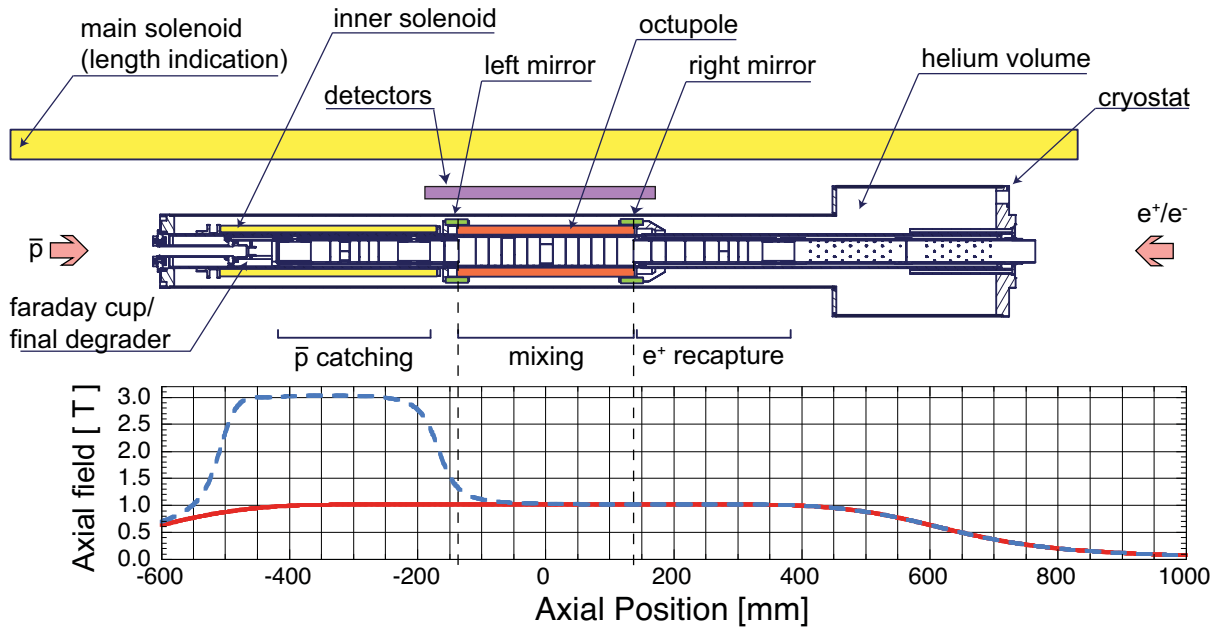}}}
\caption{(Color online) Schematic diagram of the ALPHA apparatus. Particles are confined axially in an electrostatic well formed by biasing cryogenically-cooled, cylindrical electrodes centered on the trap axis. The axial magnetic field, graphed below the schematic, confines the particles radially.  The \pbar's were caught with the inner solenoid on, in a field of 3\us T, as shown by the blue dashed curve.  The inner solenoid was ramped off before transfer of the \pbar's to the mixing region.  The experiments described here were done in the $\sim 1\,$T field shown by the red solid curve.  The MCP/Phosphor screen used to take images \cite{andr:08} of the inner regions of the \elec\ plasmas and \pbar\ clouds is located to the right of the parts of the apparatus shown here, in a field of 0.024\us T.}
\label{apparatus}
\end{figure*}

Minimum-B traps are shallow (of order 0.7\us K/T per Bohr magneton), and experimentalists have not yet learned to synthesize \Hbar\ with sufficiently low energy to be trapped.  One obstacle to progress has been the lack of detailed information about the \pbar\ cloud \cite{IsPlasma} dimensions.  Until recently, only two techniques that measure the \pbar\ radial profile have been reported in detail.  The first, based on \pbar\ annihilation on the background gas \cite{fuji:04}, yields a crude [$\sim 4\,$mm ($1\sigma$)] three-dimensional image of the \pbar\ cloud.  To observe a sufficient number of annihilations, the background gas pressure must be much higher than is normally used when synthesizing antihydrogen atoms.  This may influence the \pbar\ cloud dimensions.  The second interpolates the density profile from two destructive measurements \cite{oxle:04}: the total \pbar\ number, and the number that are located within a fixed radius set by an aperture.  The reconstruction makes assumptions about the applicability of the global thermal equilibrium state of these plasmas \cite{onei:79,dris:88}, and about the \pbar\ temperature.  We note that with our diagnostics (reported here and in \cite{andr:08}) we have seen many long-lived radial profiles that are not in global thermal equilibrium.

Recently we described a diagnostic that gives high quality information about the radial profile.  The diagnostic is based on a MCP-phosphor screen system \cite{andr:08}.  (A similar system has also been reported by the ASACUSA collaboration \cite{yama:01a}.)  Unfortunately, apertures limit the size of the \pbar\ cloud that we can measure with our MCP-phosphor system; typically we cannot measure the profile beyond radii of $1.5$--$3.0\,$mm, depending on the local magnetic field in which the \pbar's are trapped.  Some \pbar\ clouds are completely imaged by this system, but others are far larger, and can extend all the way out to the walls of our trap at radius $R_w=22.3\,$mm.  Here, using the ALPHA collaboration trap \cite{andr:07}, we describe a new diagnostic that probes the outer radial profile based on measurements of ballistic \cite{faja:05} losses induced by an octupole magnet.  After a brief description of how we load particles into the trap, we describe the diagnostic.  Then we discuss tests used to validate its performance, and close with several examples illustrating its use.

\section{Trap Loading Cycle}
We load our trap by accepting a pulse of \pbar's from the AD.  The \pbar's enter the apparatus from the left (see Fig.~\ref{apparatus}), and are slowed in a degrading foil.  They reflect from a repelling potential at the far end of the ``catching'' region of the trap, and are then captured into an electrostatic well by quickly erecting an electrostatic barrier, at the near end of the trap, before they can escape back to the degrading foil.  The \pbar's are cooled by collisions with a pre-existing electron (\elec) plasma \cite{gabr:89}.  Multiple \pbar\ pulses can be caught and cooled, each adding about 40,000 \pbar's to the trap. Typically we use four such ``stacks'' in the data presented here.  The \elec\ plasma is then ejected by fast manipulations of the electrostatic well that leave the massive \pbar's behind.    After cooling and \elec\ ejection, the \pbar's are transferred, via manipulations of the electrostatic well potentials, to the ``mixing'' region of the trap.  The octupole magnet \cite{bert:06} we use to determine the \pbar\ radial profile is centered over this region.  Positrons, when needed, are transferred from our Positron Accumulator \cite{murp:92,jorg:01} and recaptured in the region indicated in Fig.~\ref{apparatus}.  They are then transferred to the mixing region via manipulations of the electrostatic well potentials.

\section{Diagnostic Description}

\begin{figure}[t]
\centering
\centerline{\resizebox{8cm}{!}{\includegraphics{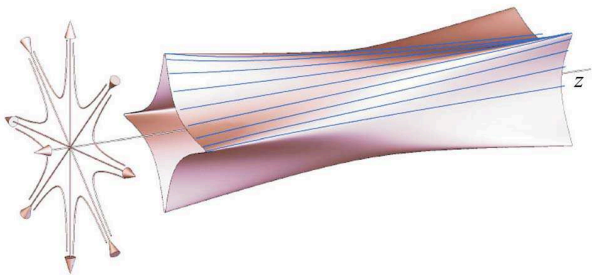}}}
\caption{(Color online) \label{field_lines}Magnetic field from the octupole and solenoid coils.  The vectors on the left represent the directions of the axially-invariant field from these coils.  The surface is created by following the field lines from a radially centered circular locus; the lines shown within the surface are field lines.}
\end{figure}

\begin{figure}[b]
\centering
\centerline{\resizebox{8cm}{!}{\includegraphics{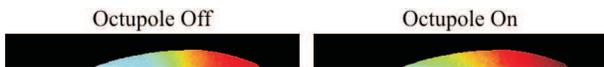}}}
\caption{(Color online) \label{Oct_Quad} Field lines imaged by passing a circular \elec plasma through the octupole with the octupole off and on.    Apertures \cite{andr:08} form the image boundaries and limit us to viewing only one quadrant of the octupole field map.  The distortion evident in the right-hand image corresponds to one of the flutes at the end of the magnetic surface shown in Fig.~\ref{field_lines}.}
\end{figure}

To understand how the radial diagnostic works, it is helpful to visualize the field lines from the solenoid and octupole coils. The field lines originating from a circular locus of points in the plane transverse to $\hat z$ form four-fluted cylindrical surfaces; the flutes at each end are rotated by $45^\circ$ with respect to each other.  An example of the resulting surfaces is shown in Fig.~\ref{field_lines}.  Fig.~\ref{Oct_Quad} shows an image of one quadrant of the field lines, generated by passing \elec's through the octupole and onto our MCP/Phosphor screen \cite{andr:08}.

\begin{figure}[t]
\centering
\centerline{\resizebox{8.5cm}{!}{\includegraphics{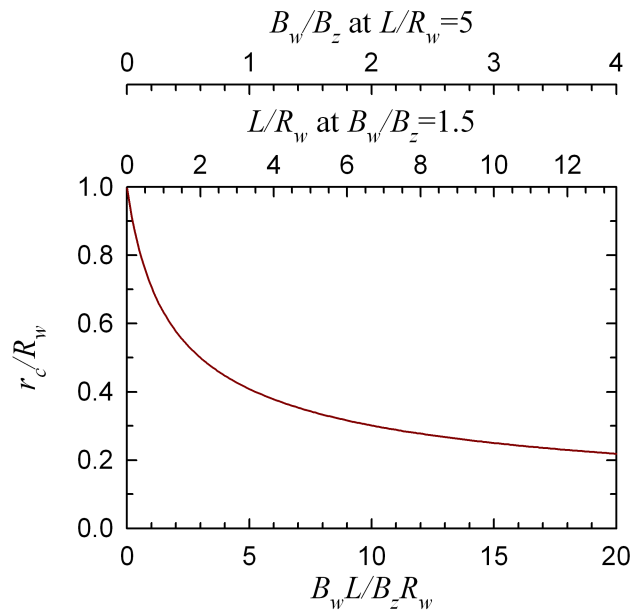}}}
\caption{(Color online) \label{Crit_Rad} The normalized critical radius [Eq.~(\ref{oct_crit})] as a function of the octupole strength $B_w$ and orbit length $L$.  The alternate axes shown at the top isolate the dependence on each parameter while holding the other fixed at a typical value.}
\end{figure}

Antiprotons confined by the electrostatic well within the octupole bounce back and forth while following the magnetic field lines \cite{FollowFieldLines}. Antiprotons that are on field lines that extend to the physical trap wall before reaching one of the electrostatic walls will follow them there and annihilate.  For a given end-to-end bounce length $L$, field lines lying outside of a critical radius $r_c$ at the trap center will hit the wall, while those lying inside the critical radius will not.  The normalized critical radius is \cite{faja:06,faja:08}:
\begin{equation}
\label{oct_crit}
\frac{r_c}{R_w}=\frac{1}{\sqrt{1+\frac{B_w}{B_z}\frac{L}{R_w}}}.
\end{equation}
This relation is depicted in Fig.~\ref{Crit_Rad}.  The longer the trap, and the stronger the octupole field, the smaller the critical radius. The normalized critical radius is never very small because the octupole field, which scales as $r^3/R_w^3$, is very weak near the trap axis relative to its strength at the wall.  This is advantageous for confinement \cite{faja:04}, as a large cloud survives and the inner core of the \pbar\ cloud is not strongly perturbed by the multipole field. However, as we show below, it limits the observable minimum \pbar\ radius to about 7\us mm for a 135\us mm long well.  If we had used a quadrupole instead of an octupole, we could have measured radial distributions to much smaller radii; for instance, to 0.24\us mm for equivalent parameters. Such a small critical radius would be very useful as a diagnostic, but could make it difficult to synthesize \Hbar.

\begin{figure}[t]
\centering
\centerline{\resizebox{8.5cm}{!}{\includegraphics{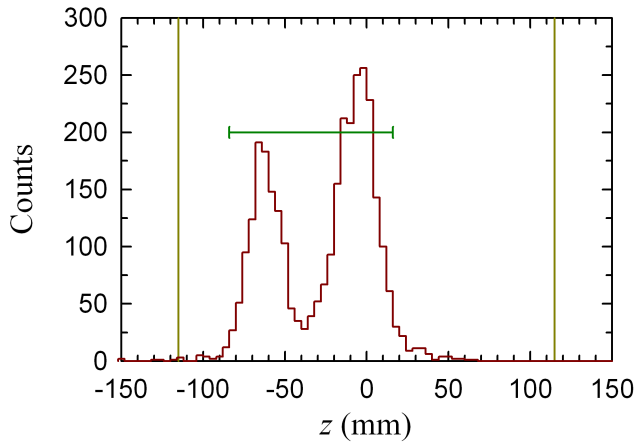}}}
\caption{(Color online) \label{z_loss} Axial positions where the \pbar's hit the trap wall under the influence of the octupole. The horizontal bar indicates the the axial extent and position of the electrostatic well confining the \pbar's. The loss is greatest near the ends of the confining electrostatic well.  The positions are determined by a position-sensitive particle detector which monitors the \pbar\ annihilation products; the vertical lines at $z=\pm 115\,$mm indicate the axial extent and position of the detector.}
\end{figure}

The ballistic loss of particles on trap walls in the presence of a multipole field was first identified with electrons in a quadrupole magnet \cite{faja:05}.  This process is easier to study with \pbar's than with \elec's, however, because individual \pbar\ annihilations can be detected and localized on the trap wall with a position sensitive detector.  The detector \cite{fuji:05} comprises three layers of silicon cylindrically arrayed around the trap axis just outside of the octupole magnet (see Fig.~\ref{apparatus}). It is not yet fully deployed, but, using a partial system consisting of 10\% of the full system, we observe (Fig.~\ref{z_loss}) that \pbar's hit the wall at the ends of the electrostatic well.  We expect to observe this type of loss pattern as it is at the ends of the trap that the accessible field lines extend furthest outward; we note, however, that annihilations tend to occur at the ends of the electrostatic well even in the absence of an octupole field \cite{fuji:04}.

For the experiments reported in Fig.~\ref{Long_Short}--\ref{WithPositrons}, annihilations were detected by scintillators coupled to Avalanche Photo Diodes (APDs). As with the silicon detector, the scintillators are cylindrically arrayed around the trap axis just outside of the octupole magnet.  Annihilations are identified by the firing of more than one scintillator in a 150\us ns coincidence window, and we detect annihilations with greater than 50\% efficiency. The detector background noise is of order a few events per second.  Timing modules correlate annihilations with experimental operations and conditions such as the strength of the octupole field.

To measure the size of a \pbar\ cloud, we first transfer it into an electrostatic well in the octupole field region; the octupole field is turned off during the transfer.  We then measure the \pbar\ kinetic energy by monitoring the rate at which the \pbar's escape as we slowly lower one endwall of the electrostatic well \cite{amor:04b}.   Typically we find that the energy is between 1 and 15\us eV; the energy depends on the details of the transfer process and the electrostatic well potentials. This measurement is destructive, but since the energy is largely set by the electrostatics, not by the \pbar\ radial profile, it is sufficient to measure this energy once for a series of profile measurements.  From this energy, we determine the bounce length $L$ of the \pbar's in the electrostatic well. The uncertainty (and spread) of the \pbar\ energy sets the uncertainty in the orbit lengths  quoted in the figure captions.  Finally, for each \pbar\ cloud that we want to analyze, we slowly ramp up the octupole field $B_w$ while monitoring the losses. From the time history of the losses, we can invert Eq.~(\ref{oct_crit}) to reconstruct the radial distribution of \pbar's:
\begin{equation}
\label{reconstruct}
n(r_c[B_w(t)])=\frac{N(t)}{2\pi r_c[B_w(t)]{\frac{dr_c}{dB_w}}{\frac{dB_w}{dt}}\Delta t}.
\end{equation}
Here $B_w(t)$ is the octupole field at time $t$, $r_c[B_w(t)]$ is the instantaneous critical radius, and $dr_c/dB_w$ is evaluated at the instantaneous field $B_w$.  The raw data from our detector is binned in intervals of time $\Delta t_0=1\,$ms; we rebin the data into intervals ranging between $\Delta t=0.333\,$s (45\us s and shorter octupole ramp times) and 1.332\us s (180\us s ramp times) to decrease the scatter.  $N(t)$ is the number of counts in the bin centered around $t$.  The mapping defined by Eqs.~(\ref{oct_crit}) and (\ref{reconstruct}) is nonlinear; points are closer together in $r$ at small radii than at large.  To further reduce the scatter at small $r$ we rebin $n(r)$ so that the spacing between successive points in $r$ is never less than 0.075\us mm.

\section{Validation Tests}

\begin{figure}[btp]
\centering
\centerline{\resizebox{8.5cm}{!}{\includegraphics{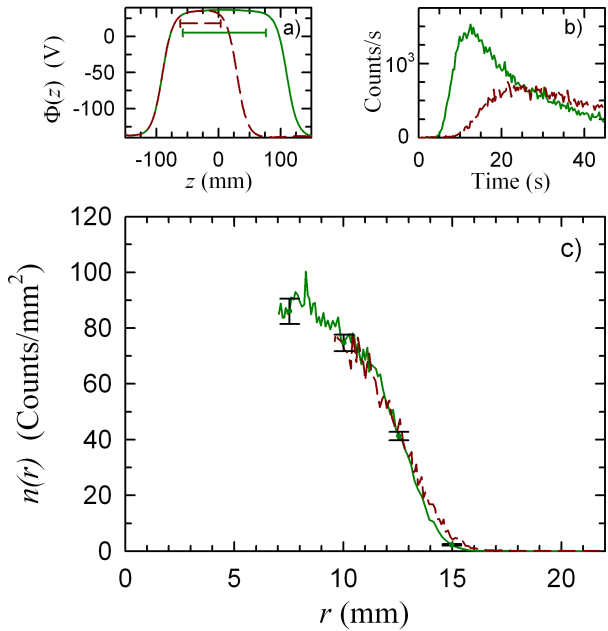}}}
\caption{(Color online) \label{Long_Short} Comparison of the radial profiles of otherwise identical \pbar\ clouds held in wells of different length. Panel~a) shows the electrostatic well potentials $\Phi(z)$ for the two cases; the horizontal bars indicate the axial extent and position of the \pbar\ orbits before the application of the octupole field.  Panel~b) shows the time history of the \pbar\ annihilations as the octupole field is ramped up. Panel~c) shows the resulting radial profiles. In all graphs, the green solid curve corresponds to the longer well ($135\pm 5\,$mm) and the red dash curve corresponds to the shorter well ($65\pm 5\,$mm).  The maximum $B_w$ at the end of the 45\us s ramp was 1.54\us T, and $B_z=1.03\,$ T. At the inner radii, Eq.~(\ref{oct_crit}) predicts that the $\pm 5\,$mm length uncertainty/spread engenders a radial uncertainty of about $\pm 0.12\,$mm at $135\,$mm, and $\pm 0.30\,$mm at $65\,$mm.  Near the wall, the uncertainty predicted by Eq.~(\ref{oct_crit}) diminishes, but the time binning engenders an uncertainty of about $\pm 0.25\,$mm. The error bars indicate the size of the typical calculated statistical error.  Both \pbar\ clouds were collected with four stacks. }
\end{figure}

Typical data are displayed in Fig.~\ref{Long_Short}, which shows the radial profile of two otherwise identically prepared \pbar\ clouds stored in wells of different length. Changing the well length should not change the radial profile of identically prepared \pbar\ clouds, and as expected, the measured profiles are almost identical over their common range. However, as predicted by Eq.~(\ref{oct_crit}), changing the well length does change the minimum radius observable with the diagnostic from about 7.0\us mm for the 135\us mm well, to 9.6\us mm for the 65\us mm well.

\begin{figure}[tbp]
\centering
\centerline{\resizebox{8.5cm}{!}{\includegraphics{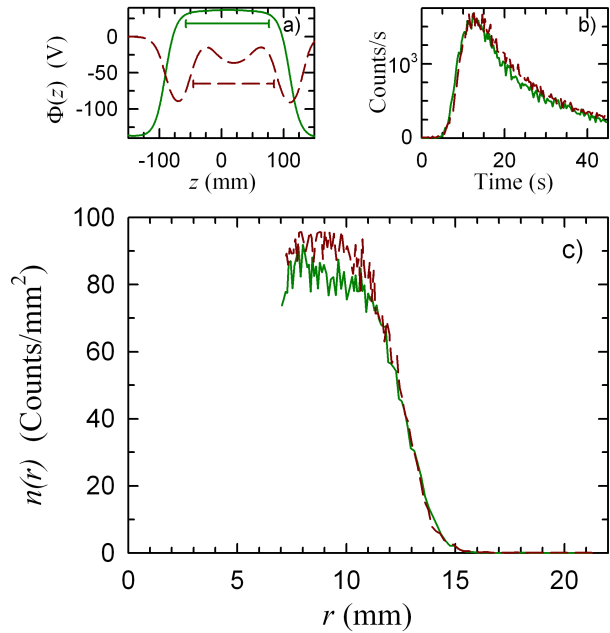}}}
\caption{(Color online) \label{Flat_Nested} Comparison of the radial profiles obtained with flat (green solid) and nested well potentials (red dash). The well lengths were $135\pm 5$ and $130\pm 5\,$mm respectively.  The graph descriptions and all other parameters are the same as in Fig.~\ref{Long_Short}. }
\end{figure}

Figure~\ref{Flat_Nested} compares the radial profiles of identically prepared \pbar\ clouds held in a flat-bottomed well, and in a nested well similar to those used to synthesize \Hbar\ \cite{amor:02}. The well length inferred from the measured \pbar\ energies was 130\us mm for the nested well, which is slightly shorter than the 135\us mm length inferred for the flat well. Changing the well shape should not change the radial profile because the azimuthally-symmetric electrostatic well fields do not induce radial transport. As expected, the measured profiles are nearly identical.  Thus, the diagnostic is indeed independent of the well shape so long as the proper well length is employed in the analysis.

\begin{figure}[btp]
\centering
\centerline{\resizebox{8.5cm}{!}{\includegraphics{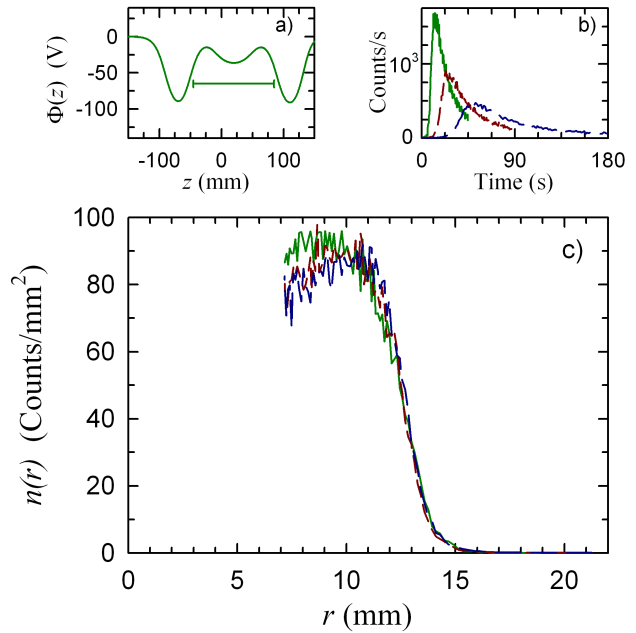}}}
\caption{(Color online) \label{Ramp_Time} Comparison of the radial profiles with octupole ramps of 45 (green solid), 90 (red short-dash), and 180\us s (blue long-dash).  The well length in each case was $130\pm 5\,$mm.  The graph descriptions and all other parameters are the same as in Fig.~\ref{Long_Short}. }
\end{figure}

As the octupole ramps, outward diffusion \cite{gils:03,faja:05} increases for those \pbar's that are still within the critical radius; if this diffusion were too fast, the profiles would be suspect.  We have established that the diffusion is not fast on the time scale of the octupole ramp by comparing (Fig.~\ref{Ramp_Time}) the radial profiles of identically prepared \pbar\ clouds taken with ramps of 45 (our standard ramp), 90, and 180\us s.  The differences between the curves are not large.

\begin{figure}[tbp]
\centering
\centerline{\resizebox{8.5cm}{!}{\includegraphics{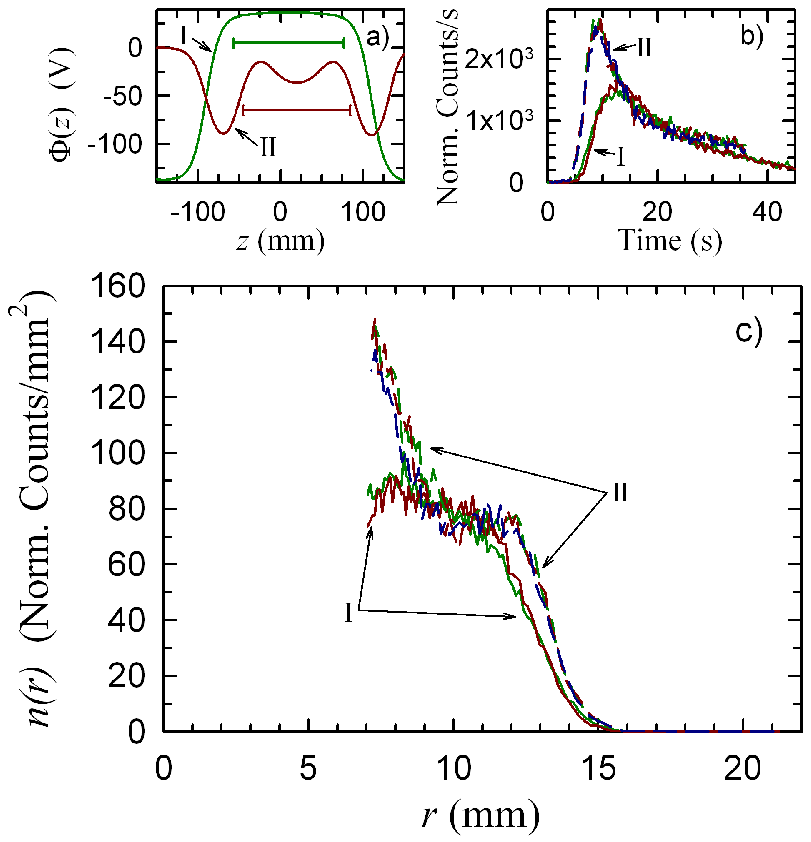}}}
\caption{(Color online) \label{ShotToShot} Radial profiles for two sets of \pbar\ clouds that were prepared differently; the \elec\ cooling plasmas used for the two sets came from different \elec\ sources.   The figure also shows that the load-to-load reproducibility of the \pbar\ clouds is high;  the set labeled I compares two loads, while the set labeled II compares three. The AD and our apparatus can be quite reproducible; the two profiles in set I were measured 23 hours apart on different AD shifts.  (Note that the clouds were analyzed in different shape wells, one (I) of length $135\pm 5\,$mm in a flat well, and the other (II) of length $130\pm 5\,$mm in a nested well. The ramp time for set II was slightly shorter than for set I: 36\us s instead of 45\us s.  However, as verified in Figs.~\ref{Flat_Nested} and \ref{Ramp_Time}, these difference should not affect the radial analysis. Finally, only two stacks were used in set II; the profiles for this set were normalized to four stacks.)  The graph descriptions and all other parameters are the same as in Fig.~\ref{Long_Short}.}
\end{figure}

The diagnostic described here would have little utility if all reconstructed radial profiles were identical; Figure~\ref{ShotToShot} shows that radial profiles of \pbar\ clouds that are differently prepared can be dissimilar. Figure~\ref{ShotToShot} also shows that the load-to-load reproducibility of the \pbar\ profiles is quite good.

Measurements taken with our MCP/phosphor screen diagnostic confirm that the central density is not significantly perturbed by cycling the octupole field.  For instance, for parameters identical to the nested well profile shown in Fig.~\ref{Flat_Nested}, the total number of \pbar's within the MCP/phosphor apertures varied by less than 4\% on two successive shots, one with the octupole off and one with it ramped up and then back down.    This discrepancy is well within the shot-to-shot variation of our loads.  This result, taken together with the results shown in Figs.~\ref{Long_Short}--\ref{Ramp_Time}, establish that ramping the octupole field is a robust method of obtaining the radial profile that is largely independent of the details of the ramp speed and well shape.

\section{Observations}

We have used our new diagnostic to characterize our \pbar\ manipulation sequences, and to study interesting physics issues.  In this section, we outline four of these measurements; all need further study.

\begin{figure}[btp]
\centering
\centerline{\resizebox{8.5cm}{!}{\includegraphics{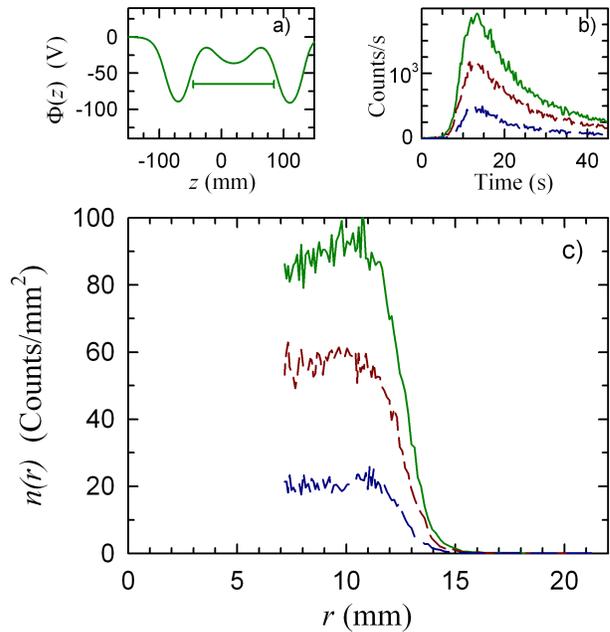}}}
\caption{(Color online) \label{Stacks} Comparison of the radial profiles for two (blue long-dash), three (red short-dash), and four (green solid) stacks.  The well length was $130\pm 5\,$mm.  The graph descriptions and all other parameters are the same as in Fig.~\ref{Long_Short}. }
\end{figure}

As described earlier, we can stack multiple \pbar\ pulses from the AD.  Figure~\ref{Stacks} shows the \pbar\ profile for two, three, and four stacks.  The stacks add to each other without significantly changing the radial profile.  The results obtained when only one stack is accumulated are quite different, however.  The profile is completely contained within a radius of 7\us mm and is not visible with this diagnostic.  We suspect that the difference is due to straggler \elec's from the degrader accidentally captured during the first (and subsequent) \pbar\ injections. These \elec's are captured by the same electrostatic well manipulations used to capture the \pbar's. After capture, they cool and thermalize via cyclotron radiation and collisions, and join the deliberately captured cooling \elec\ plasma; we observe that the number of \elec's in this plasma increases with the number of stacks.  The straggler \elec's are likely emitted from the degrader over the entire area hit by the \pbar's, and, if the radius of this area is greater than the radius of the deliberately injected \elec\ plasma, the plasma radius will increase.  This will increase the size of the captured \pbar\ cloud \cite{andr:08}.  It will also increase the fraction of the degraded \pbar's captured \cite{andr:08}; we observe this fraction increasing from about 45\% on the first stack to over 90\% on later stacks.

\begin{figure}[btp]
\centering
\centerline{\resizebox{8.5cm}{!}{\includegraphics{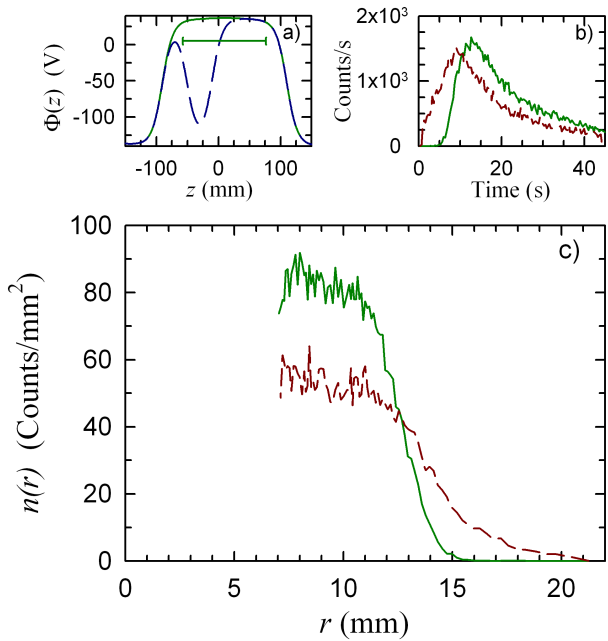}}}
\caption{(Color online) \label{SingleMany} Comparison of the radial profiles obtained with a gentle (green, solid) and abrupt (red short-dash) injection into a long well.  The blue dash curve in Panel~a) shows the pre-injection well structure (the \pbar's start in the leftmost well) and the green solid curve shows the final well, which has a length $135\pm 5\,$mm. The graph descriptions and all other parameters are the same as in Fig.~\ref{Long_Short}. }
\end{figure}

The transfer process from the catching region of our trap to the mixing region leaves the \pbar's situated in a short well on one side of the final trapping well. From this short well, the \pbar's are injected into the final well.  Normally, we do this gradually, by smoothly changing the potentials over a 1\us ms time period.  When we change the potentials abruptly, on a time scale of approximately $3\,\mu$s, \pbar's are lost on injection, and the \pbar\ cloud's radius increases significantly, as shown in Fig.~\ref{SingleMany}.  There is no obvious mechanism for the immediate loss and cloud expansion.

\begin{figure}[tbp]
\centering
\centerline{\resizebox{8.5cm}{!}{\includegraphics{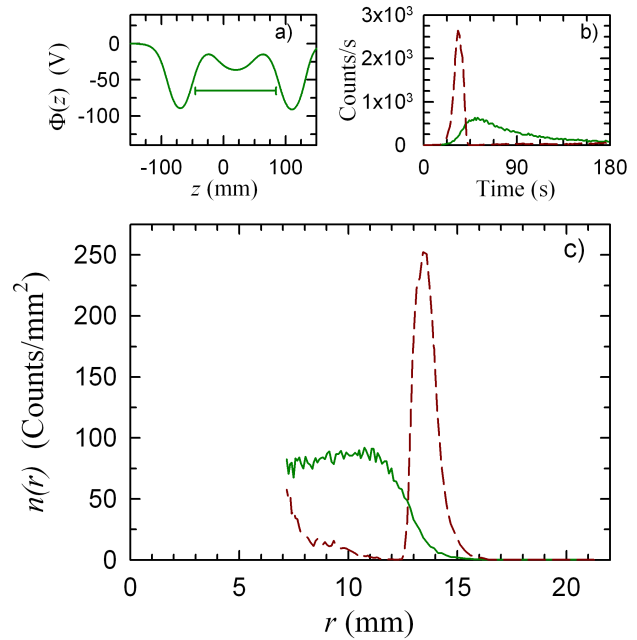}}}
\caption{(Color online) \label{Centrifugal} Comparison of the radial profiles with and without electrons (dashed red and solid green lines, respectively.  The well length was $130\pm 5\,$mm, and the ramp time was $180\pm 5\,$s. The graph descriptions and all other parameters are the same as in Fig.~\ref{Long_Short}. }
\end{figure}

Figure~\ref{Centrifugal} shows the very different radial profile obtained when we do not eject the \elec's before transfer and analysis.  The antiprotons form a hollow ring around the trap center.  This type of distribution is compatible with the global thermal equilibrium of a mixed \elec-\pbar\ plasma, which places the \pbar's in a halo surrounding the \elec\ plasma \cite{onei:81} when the particles are sufficiently cold.  However, we observed losses during the transfer process that could have preferentially hollowed the distribution and produced the observed profile.

Note that the \pbar's likely cool via collisions with the \elec's during the octupole ramp.  This would shorten the axial extent of the \pbar\ orbits, and thus introduce some uncertainty into the reconstruction of the radial profiles via Eq.~(\ref{reconstruct}) as it introduces variation in $L$.  This is particularly true if the \pbar's cool into the side wells, where their orbit length would decrease abruptly by more than a factor of two.  This effect would cause us to erroneously reconstruct, via Eq.~(\ref{reconstruct}), some charge to be at falsely low radii, probably below the 7\us mm radius visible to us with this diagnostic.  Thus, cooling does not explain the halo visible in Fig.~\ref{Centrifugal}.  This very interesting result needs further study.

\begin{figure}[tph]
\centering
\centerline{\resizebox{8.5cm}{!}{\includegraphics{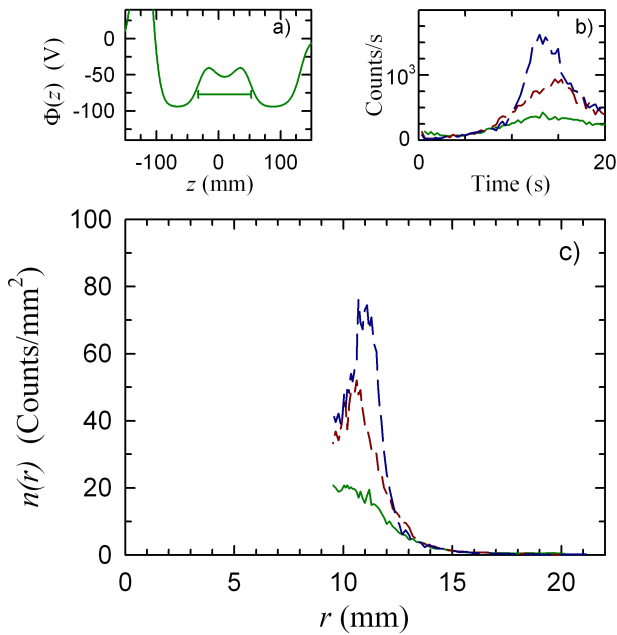}}}
\caption{(Color online) \label{WithPositrons} Comparison of the \pbar\ radial profile with different density positron plasmas.  The green, solid curve shows the profile with no \pos, and the red short-dash and blue long-dash curves show the profile with 13\us million and 25\us million \pos\ respectively.  The well length was $85\pm 5\,$mm, the maximum field was 1.20\us T, and the ramp time was 20\us s. Only one stack was captured, but the \elec\ cooling plasma was created by a secondary \elec\ source which makes a large radius plasma; thus, unlike in Fig.~\ref{Stacks}, \pbar's are visible with only one stack. The graph descriptions and all other parameters are the same as in Fig.~\ref{Long_Short}. }
\end{figure}

Finally, in Fig.~\ref{WithPositrons}, we show radial profiles for a mixed \pos-\pbar\ plasma.  As the density of the \pos\ plasma is increased, \pbar's appear to be transported outward.  Here, as described in the previous paragraph, the interpretation of the results is complicated by cooling of the \pbar's (on the \pos\ in this case.)  Cooling will again cause some charge to appear at falsely low radii, and this very likely causes us to underestimate the outward movement of the \pbar's.

A possible explanation of the outward movement shown in Fig.~\ref{WithPositrons} is that it is the result of the formation of highly excited \Hbar\ that is either 1) ionized at the radial edge of the \pos\ plasma by its self consistent electric field, which is strongest at the edge, or 2) ionized by the vacuum electrostatic well fields.  Note that the \pbar's from \Hbar\ that was ionized within the \pos\ plasma radius would have the opportunity to recombine into \Hbar\ again, while those at larger radii would orbit unperturbed.  With time, the \pbar's remaining in the \pos\ plasma would be swept out to larger~radii.  Unpublished simulations of realistic antihydrogen formation/field ionization cycles, using the code described in \cite{robi:04}, found similar transport.  We do not yet have any other direct experimental evidence that this cycling is occurring.

\section{Conclusions}

We have shown that we can determine the outer radial profile of \pbar's stored in a Penning-Malmberg trap by monitoring the losses induced by ramping an octupole magnet.  This technique complements direct imaging of the inner radial profile \cite{andr:08}, and provides more precise and reliable information than earlier techniques \cite{fuji:04,oxle:04}.  We have tested the diagnostic by varying the electrostatic well length and shape, and by varying the ramp time, and we have used the diagnostic to study several procedures and manipulations pertinent to the synthesis of antihydrogen atoms.

This work was supported by CNPq, FINEP (Brazil), ISF (Israel), MEXT (Japan), FNU (Denmark), NSERC, NRC/TRIUMF (Canada), DOE (USA), EPSRC and the Leverhulme Trust (UK) and HELEN/ALFA-EC.


\begin{thebibliography}{27}
\bibitem[{IsP()}]{IsPlasma}
\bibinfo{note}{Depending on the \pbar\ density and temperature, the \pbar\
  cloud may be in the single particle regime, the plasma regime, or in between
  the two. We will refer to it as a cloud throughout this paper.}
\bibitem[{Fol()}]{FollowFieldLines}
\bibinfo{note}{Since the magnetic field is strong ($>1\,$T), the \pbar's are,
  apart from a slow $\mathbf{E}\times\mathbf{B}$ rotation around the trap axis,
  tightly bound to the field lines.}
\end{thebibliography}

\end{document}